\documentstyle[12pt]{article}\textheight 240mm\textwidth 160mm
\renewcommand{\baselinestretch}{1.5}

\begin{document}
\hspace*{11cm}MPI-Ph/92-40\\
\hspace*{11.6cm}June 1992\\
\hspace*{11.6cm}DPKU-9212
\vspace*{3mm}
\begin{center}{\bf A new insight into BRST anomalies in string theory}
\end{center}
\vspace*{2mm}
\begin{center}{\sc Takanori Fujiwara}\\
{\sl Department of Physics, Ibaraki University, Mito 310 (Japan)}\\
{\sc Yuji Igarashi}\\ {\sl Faculty of General Education, Niigata University,
Niigata 950-21 (Japan)}\\
{\sc Jisuke Kubo}$^{*}$\\
{\sl Max-Planck-Institut f\"ur Physik\\
$-$ Werner-Heisenberg-Institut $-$\\
P.O.Box 40 12 12, Munich (Fed. Rep. Germany)}\\
and\\
{\sc Kayoko Maeda}\\ {\sl Department of Physics, Niigata University,
Niigata 950-21 (Japan)}
\end{center}
\vspace*{1mm}

\begin{center}
ABSTRACT
\end{center}
$\ \ \ $Using the generalized hamiltonian method of Batalin,
Fradkin and Vilkovisky, we investigate the algebraic structure of
anomalies in the Polyakov string theory
that appear as the Schwinger terms
in super-commutation relations between BRST charge and total hamiltonian.
We obtain the
most general form of the anomalies in the extended phase space,
without any reference to a two dimensional metric.
This pregeometrical result, refered to as the genelarized Virasoro
 anomaly,
independent of the gauge and the regularization under a minor
assumption, is a non-perturbative result,
and valid for any space-time dimension.
In a configuration space, in which the two dimensional metric
can be identified,
we can geometrize the result without assuming the weak
gravitational field, showing that
the most general anomaly exactly exhibits the Weyl anomaly.

\vspace*{4mm}
\footnoterule
\vspace*{3mm}
\small{$^{*}$Permanent address: College of Liberal Arts, Kanazawa University,
Kanazawa 920, Japan}
%\end{document}
\newpage

\pagestyle{plain}
\hoffset=-1.5cm
\voffset=-1.5cm
\renewcommand{\baselinestretch}{1.5}

%\begin{document}
\newcommand{\bi}{\bibitem}
\newcommand{\be}{\begin{eqnarray}}
\newcommand{\ee}{\end{eqnarray}}
\newcommand{\nn}{\nonumber}
\newcommand{\al}{\alpha}
\newcommand{\bt}{\beta}
\newcommand{\ga}{\gamma}
\newcommand{\de}{\delta}
\newcommand{\ep}{\epsilon}
\newcommand{\ve}{\varepsilon}

\newcommand{\vc}{\vartheta}

\newcommand{\ka}{\kappa}
\newcommand{\la}{\lambda}

\newcommand{\x}{\xi}
\newcommand{\p}{\pi}
\newcommand{\r}{\rho}
\newcommand{\si}{\sigma}

\newcommand{\vp}{\varphi}
\newcommand\Ga{\Gamma} \newcommand\De{\Delta}
\newcommand\La{\Lambda}

\newcommand{\dl}{\partial}
\newcommand{\cd}{\cdot}
\newcommand{\Dm}{\cal D}
\newcommand{\cH}{\cal H}
%
%References
\newcommand{\Ram}{Ann. Math. }
\newcommand{\Rap}{Ann. Phys. (N.Y.) }
\newcommand{\Rcmp}{Comun. Math. Phys.}
\newcommand{\Rhpa}{Helv. Phys. Acta }
\newcommand{\Rmpl}{Mod. Phys. Lett. }
\newcommand{\Rnc}{Nuovo Cimemto }
\newcommand{\Rnp}{Nucl. Phys. }
\newcommand{\Rpl}{Phys. Lett. }
\newcommand{\Rplc}{Phys. Rep. }
\newcommand{\Rpr}{Phys. Rev. }
\newcommand{\Rprl}{Phys. Rev. Lett. }
\newcommand{\Rprs}{Proc. Roy. Sci. }
\newcommand{\Rptp}{Prog. Theor. Phys. }
\newcommand{\Rtmp}{Theor. Math. Phys. }
\newcommand{\Rzsp}{Z. Phys. }
\newcommand{\Rrnc}{Riv. Nuovo Cimento }
%
%%%%%%
\newpage
%%%%%%%%%%%%
\section{Introduction}
It is well known that in quantizing relativistic strings
at subcritical dimensions
one encounters anomalies that appear
in different forms;
the central extension of the Virasoro algebra,
the Weyl anomaly,
the non-vanishing square of BRST
charge etc.\footnote{See
ref. \cite{GSW} for a review and references therein.}.
Most of the results
on these anomalies,
as those in the classical references such as refs. \cite{FV}-\cite{Fr},
however, are obtained only in particular classes of gauges. A general
consideration, in which  the gauge dependence of these anomalies is
satisfactorily explored, has been lacking, and
this may be one of the reasons why the relations among the
anomalies above have been
only partially revealed \cite{PolA}-\cite{FISA}.

The phase space of Batalin, Fradkin and Vilkovisky (BFV)
\cite{FVA}
is much lager than the usual
one, and the gauge-dependence in their hamiltonian formalism
can be tracked in a very transparent manner. Therefore,
one may expect to
obtain the most general form of anomalies
in the extended phase space and then
to gain a unified understanding on the anomalies \cite{FIKA}.
Showing this
in the Polyakov string theory \cite{PolA} is exactly the aim of the present
paper. We shall perform an algebraic
analysis of anomalies \cite{WZA,BRSA,pr}
in the extended phase space \cite{FVA} of the theory
in an exhaustive fashion, and we
will get
a new insight into the anomalies in string theory by revealing
their pregeometrical origin and hierarchical relationships.

As we will explain in sect. 2,
there is a general, simple criterion for the presence of
a gauge anomaly;
given a gauge invariant classical system,
the anomaly exists if the classical gauge algebra of
BRST charge $Q$ and total hamiltonian $H_{\rm T}$ can not
be maintained upon quantization, i.e. if
$2 Q^2 = [Q\ ,\ Q] \ \neq\  0$ and/or
$[Q\ ,\ H_{\rm T}] \neq 0.$
We impose super-Jacobi identities
\cite{Man,KaMa,FIKA}\cite{ff}-\cite{fik}
on these anomalous commutators
and expand them in $\hbar$ \cite{FIKA}.
This leads
to a set of consistency conditions  on the anomalous
terms \cite{WZA}
at each order in $\hbar$,
which in turn exhibits the descending nature \cite{StrA}
of the anomalous Schwinger terms for $[Q\ ,\ Q]$ and
$[Q\ ,\ H_{\rm T}] $ in the hamiltonian formalism
\cite{FIKA,fik,IKS}.
The fact that in the BFV formalism the BRST charge
is directly constructed from classical
first-class constraints \cite{FVA} implies
that
the algebraic structure of $Q^2$ can be investigated in a
completely gauge-independent fashion \cite{FIKA}.
The gauge dependence of the theory enters only as a BRST commutator
in the total hamiltonian in the extended phase space, and
can be easily tracked in the BFV formalisms.

In sect. 3, we shall
solve all the consistency conditions
in the full expended phase space
in an exhaustive manner,
under the assumption specified there.
Since the
BRST cohomology in the extended phase space
is the most general one, so is the solution, too.
This is the anomaly in the BFV formulation
of the Polyakov string theory, and is a pregeometrical result because it is
obtained without any reference to a metric.

This pregeometrical anomaly, called there the generalized
Virasoro anomaly with its very simple form, can not be directly
identified  with the reparametrization or Weyl anomaly
in the extended phase space.
To do this, we must go to the configuration space
and
geometrize our result,
as we shall do in sect. 4.
For a configuration space
in which three metric variables can be independently defined,
we find a very general result that
the generalized Virasoro anomaly
exhibits nothing but the Weyl anomaly and
completely fixes the relation among the anomalies we mentioned
at the beginning. We emphasize that the general form of the Weyl anomaly,
derived
from the generalized Virasoro anomaly, is a non-perturbative
result and independent of a choice of gauges and regularizations,
and that it is valid for any space-time dimension D.

{}From the explicit computation of the Weyl
anomaly in the conformal gauge \cite{FujG}, we fix the overall
factor for the generalized Virasoro anomaly. This enables us to
derive the $Q^{2}$ anomaly in the orthonormal gauge
\cite{KOA} including its absolute normalization,
which is the BRST version of the conventional Virasoro anomaly.
This demonstrates
one of the examples that show the hierarchical relations among the anomalies.

The appendix is devoted to prove the uniqueness of the solutions
to the consistency conditions of sect. 3.

\section{Commutator anomalies and
consistency conditions}
Here we would like to briefly  outline the basic idea in
our previous work on commutator anomalies \cite{FIKA}.
The BFV method \cite{FVA}
\footnote{See ref. \cite{HenA} for a review.} uses two fundamental objects; a
BRST charge $Q$  \cite{BRSA} and a
total hamiltonian $H_{\rm T}$. They obey at the classical level
the fundamental gauge algebras
\be
\{ Q\ ,\ Q\}&=&0\ ,
\ee
and
\be
{d\over dt}Q &= &{\dot Q}\ =\ \{Q\ ,\ H_{\rm T}\}\ =\ 0\ ,
\ee
which entirely express gauge invariance of a given theory:
The "nilpotency of $Q$" expressed in eq. (1) means that the underlying
constraints in the theory are first-class while eq. (2) implies
the consistency of the constraints with the dynamics of the system.
At the quantum level, these quantities must be suitably regularized
to become well-defined operators.
An anomaly arises if
these gauge algebras can not be maintained upon quantization.
The anomalous terms may be expanded in $\hbar$ as
\be
[Q\ ,\ Q] &\equiv & i\hbar^2\Omega+{\rm O}(\hbar^3)
\ee
\be
[ Q\ ,\ H_{\rm T}] &\equiv& {i\over2}\hbar^2\Ga+{\rm O}(\hbar^3)\ ,
\ee
where $[\ ,\ ]$ denotes super-commutator. We must distinguish
a super-commutator from
a naive one $[\ ,\ ]_0  $ which is defined via
super-Poisson bracket $\{\ ,\ \}$:
\be
[A\ ,\ B]_0 &\equiv& i\hbar \{A\ ,\ B\}\ .
\ee
Our basic assumption is that the super-commutation relations between
$Q$ and $H_{\rm T}$ obey the commutation law,
\be
[A\ ,\ B]
&=&-(-1)^{\vert A\vert\vert B\vert}[ B\ ,\ A]\ ,
\ee
the distribution law
\be
[A\ ,\ B+C]&=&[A\ ,\ B]+[A\ ,\ C]\ ,
\ee
and the super-Jacobi identity
\be
& &[A\ ,\ [B\ ,\ C]\ ]
+(-1)^{\vert C\vert(\vert A\vert+\vert B\vert)}[C\ ,\ [A\ ,\ B]\ ]\nn \\
& &+(-1)^{\vert A\vert(\vert B\vert+\vert C\vert)}
[B\ ,\ [C\ ,\ A]\ ] \ =\  0,
\ee
where $\vert A\vert$ is the grassmann parity of the operator $A$
and can be either even ($=0\ \mbox{mod}\ 2$) or odd
($=1\ \mbox{mod}\ 2$).
The crucial observation is that the outer commutators in the super-Jacobi
identities for $Q$ and $H_{\rm T}$, i.e.
\be
[Q\ ,[Q\ ,\ Q]\ ]&=&0\ ,
\ee
and
\be
 2\ [Q\ ,\ [Q\ ,\ H_{\rm T}]\ ]+[H_{\rm T}\ ,\ [Q\ ,\ Q]\ ]&=&0\ ,
\ee
define a set of consistency conditions at each order in $\hbar$ in terms of
the naive
commutators only.  Therefore,
the introduction of the naive commutators is essential
for the order-by-order investigation.
In the lowest order, one finds that \cite{FIKA}
\footnote{The consistency condition (11) has been considered
in the context of bosonic string in ref. \cite{Man}.  However,
(12) which exhibits
the descending nature of anomalies has not been discussed there (see below).}
\be
\de\Omega &= &0
\ee
\be
\de\Ga &= &\{H_{\rm T}\ ,\ \Omega\}\  =\ - \dot{\Omega}\ ,
\ee
where $\de$ is the BRST transformation given
by a naive commutator,
$\de A \equiv - \{Q\ ,\ A\} = i [Q\ ,\ A]_{0}/\hbar$.
The true anomalies $\Omega$ and $\Ga$ should be
cohomologically non-trivial;
if $\Omega$ and $\Ga$ are solutions, then $\Omega + \de X$ and
$\Ga + \{H_{\rm T}\ ,\  X\} + \de Y$
also solve (11) and (12), respectively, for any $X$ and $Y$,
which however can be removed to order $\hbar^2$ by redefining $Q$ and
$H_{\rm T}$ as $Q\rightarrow Q-(\hbar X/2)$ and  $H_{\rm T}\rightarrow
H_{\rm T} - (\hbar Y/2)$.

As one can see from the definitions
(3) and (4), the anomaly in $Q^{2}$, which is basically
the commutator anomaly in the algebras of constraints,
descends into the anomaly in current divergence.
In the chiral Yang-Mills
theory, these are the Schwinger terms in the algebra of the Gau{\ss} law
constraints \cite{ff}
and the axial current divergence, respectively.
The descending nature of the anomalies, described here in the hamiltonian
 formalism, originates from the super-Jacobi identities (9) and (10),
which can not easily be recognized
in the conventional, pure mathematical
formulation of
 the descent equations \cite{StrA}
\footnote{We regard $\Ga$ as the descendant of $\Omega$ because
$\Ga$ is basically determined by $\Omega$ via eq. (12). Note that
this order of what whose descendant is, is reverse to the usual one.}.
To our
knowledge, ref. \cite{FIKA} is the
first one which delivers such a theoretical meaning
of the descent equations.
With these general discussions in mind we next would like to consider
bosonic string theory.

\section{Solution in the extended phase space:
The generalized
Virasoro anomaly}
The Polyakov string theory can be described by the lagrangian \cite{PolA}
\be
{\cal L} &=& - {1 \over 2} \sqrt{-g}g^{\al \bt}\
{\dl}_{\al} X^{\mu} {\dl}_{\bt} X_{\mu}\  ,
\qquad (\al,\ \bt = 0,1; \quad
\mu = 0,\cdots,D-1)\ .
\ee
We choose the parametrization for
the metric variables $g_{\al \bt}$ as
\be
 {\la}^{\pm} = \frac{\sqrt{-g} \pm g_{01}}{g_{11}},
\quad \xi = \ln g_{11}.
\ee
Here, the ${\la}^{\pm}$ are Weyl invariant, and we shall see below why
these variables are taken.
A Weyl transformation is described by a translation in $\xi$-variable.
The
conjugate momenta of these variables, which we denote by $\pi^{\la}_{\pm}$ and
$\pi_{\xi}$,
vanish identically.
  These are primary constraints;
\be
\pi^{\la}_{\pm} \sim 0~ , \quad
\pi_{\xi} \sim 0~.
\ee
The Dirac algorism further leads to the secondary constraints,
the Virasoro constraints,
\be
\vp_{\pm} &\equiv& {1 \over 4} ( P \pm X^{\prime})^2\ \sim 0\ ,
\ee
where $P_{\mu}$ denotes the conjugate momentum of the string coordinate
$X^{\mu}$, and $X^{\prime} = \dl X
\equiv {\dl}_1 X$.
They satisfy the algebra under the
Poisson bracket:
\be
\{ \vp_{\pm} ( \sigma )\ ,\ \vp_{\pm} ( \sigma') \} &=& \mp
(\ \vp_{\pm} ( \sigma ) \partial' - \vp_{\pm} ( \sigma') \partial\ )\
\delta ( \sigma - \sigma') \\
\{ \vp_{\pm} ( \sigma )\ ,\ \vp_{\mp} ( \sigma')
\} &=& 0 \nn\ .
\ee

The extended phase space of BFV is defined as including
to the classical phase space the
ghost-auxiliary field sector
\be
( {\cal C}^{\rm A}\ ,\ \overline{{\cal P}}_{\rm A} )\ ,\
( {\cal P}^{\rm A}\ ,\ \overline{{\cal C}}_{\rm A} )\ ,\mbox{and}\
( N^{\rm A}\ ,\ B_{\rm A} )\ ,
\ee
where  ${\rm A} ( = {\la}^{\pm},~\xi,~\pm)$ labels the  first-class constraints
given in (15) and (16).
The $ {\cal C}^{\rm A}$ and ${\cal P}^{\rm A}$ are the BFV ghost fields
carrying one unite of the ghost number,
$\mbox{gh}({\cal C}^{\rm A}) = \mbox{gh}({\cal P}^{\rm A}) =1$,
while $~~\mbox{gh}(\overline{{\cal P}}_{\rm A}) =
\mbox{gh}(\overline{{\cal C}}_ {\rm A}) = -1$
for their canonical momenta, $\overline{{\cal P}}_{\rm A}$ and
$\overline{{\cal C}}_ {\rm A}$.
The last canonical pairs in (18) are
auxiliary fields and carry no ghost number.
We assign $0$ to the canonical dimension of $
X^{\mu}, {\la}^{\pm}$ and $\xi$,
and correspondingly $+1$ to $P_{\mu},\pi^{\la}_{\pm}$ and
$\pi_{\xi}$. The canonical
dimensions of
${\cal C}_{\la}^{\pm},~ {\cal C}^{\xi},~\overline{{\cal P}}_{\pm}$
, $~\overline{{\cal P}}^{\la}_{\pm}$ and $\overline{{\cal P}}_{\xi}$
are fixed only relative to that of ${\cal C}^{\pm},
 c~ \equiv \mbox{dim}({\cal C}^{\pm})$:
\be
\mbox{dim}({\cal C}_{\la}^{\pm})~ =~ \mbox{dim}({\cal C}^{\xi})~=~1+c~ ,~
\mbox{dim}(\overline{{\cal P}}_{\pm})~ =~ 1-c~
,~  \mbox{dim}(\overline{{\cal P}}^{\la}_{\pm})~ =~
\mbox{dim}(\overline{{\cal P}}_{\xi})~=~-c~.
\ee
The canonical
dimensions of other fields are not
needed for our purpose as we will see later.
Given the constraints with the algebra (17) and the corresponding
extended phase space (18), we can construct the BRST charge
\be
 Q &=& \int d \sigma [\ {\cal C}_{\la}^{+} {\pi}^{\la}_{+}
    + {\cal C}_{\la}^{-} {\pi}^{\la}_{-} + {\cal C}^{\xi} \pi_{\xi}
    + {\cal C}^{+} ( \vp_{+} + \overline{{\cal P}}_{+}
       \partial {\cal C}^{+} ) \nn \\
 & & + {\cal C}^{-} ( \vp_{-} -\overline{{\cal P}}_{-}
        \partial {\cal C}^{-} )
     + B_{\rm A} {\cal P}^{\rm A}\ ]\ ,
\ee
with $\mbox{gh}(Q) = 1$ and $\mbox{dim}(Q)=1+c$.
 This $Q$ generates
the BRST transformations
\be
\de X^{\mu} & =& {1 \over 2} [\ {\cal C}^{+}(P + X^{\prime})^{\mu}
              +  {\cal C}^{-} (P - X^{\prime})^{\mu}\ ] \nn \\
\de P^{\mu} & =& {1 \over 2} \dl [\ {\cal C}^{+}(P + X^{\prime})^{\mu}
              - {\cal C}^{-}(P - X^{\prime})^{\mu}\ ]  \nn\\
\de {\cal C}^{\pm} & =& \pm {\cal C}^{\pm} \dl {\cal C}^{\pm} \nn\\
\de \overline{{\cal P}}_{\pm} & =&
              - [\ \vp_{\pm} \pm 2 \overline{{\cal P}}_{\pm}
              \dl {\cal C}^{\pm} \pm \dl \overline
              {{\cal P}}_{\pm} {\cal C}^{\pm}\ ] \\
\de {\la}^{\pm} &=& {\cal C}_{\la}^{\pm}~,~\de \xi~ =~ {\cal C}^{\xi} \nn\\
\de{\pi}^{\la}_{\pm} &=&\de\pi_{\xi}\ =\  0\ ,\
\de {\cal C}_{\la}^{\pm}\ =\de {\cal C}^{\xi}\ =\ 0\nn\\
\de\overline{{\cal P}}^{\la}_{\pm} &=& -\pi^{\la}_{\pm}\ ,\
\de\overline{{\cal P}}_{\xi} = -\pi_{\xi}\ ,\ \nn\\
\de N^{A} &=& {\cal P}^{A}\ ,\ \de\overline{C}_{A}\ =\ -B_{A}\ ,\
\de {\cal P}^{A}\ =\ \de B_{A}\ =\ 0\ \nn\\
(\ A &=& {\la}^{\pm},\xi, \pm\ )\ . \nn
\ee

We are now ready to solve the consistency condition
(11) and seek the solution in the form
\be
\Omega = \int d \sigma \omega\ ,
\ee
where we assume that $\omega$ is a polynomial of local operators with
$~\mbox{gh}(\omega) = 2~$ and $~\mbox{dim}(\omega) = 3~+~2c$.
According to the general structure of the BFV formalism,
the total phase space
can be divided, with respect to the action of $\de$,
into two sectors;
\be
& &{\rm S}_{1}\ \mbox{ consisting of}\
(X^{\mu}\ ,\ P_{\mu})\ \mbox{and}\ ({\cal C}^{\pm}
,\overline{{\cal P}}_{\pm})\ ,
\ \mbox{and}\nn\\
& &
{\rm S}_{2}\ \mbox{consisting of all the other fields}\ .
\ee
It is easy to see that on each sector the $\de$ operation closes:
\be
{\de_{1}}^2 &=& {\de_{2}}^2\ =\ 0\ ,\
\de_{1} \de_{2} + \de_{2} \de_{1}\  =\ 0\ ,
\ee
where $\de =\de_{1} + \de_{2}$, and $\de_{1} (\de_{2})$ acts on ${\rm S}_{1}$
(${\rm S}_{2}$) variables only.
The ${\rm S}_{2}$-sector is BRST trivial because it is made of pairs
$(U^{a}\ ,\ V^{a})$
 with $\de_{2} U^{a}= \pm V^{a}$.
As shown in ref. \cite{FIKA}, there exists no non-trivial solution to
$\de \Omega =\int d \sigma\de \omega= 0$ if $\omega$
contains the ${\rm S}_{2}$-variables.
Therefore, $\omega$ is a function
of ${\cal C}^{\pm},\overline{{\cal P}}_{\pm},X^{\mu},
P_{\mu}$ and their derivatives
only.

To proceed, we
note that the BRST charge given in (20) has rigid symmetries;
it is  a Lorentz scalar,
invariant under the translation, $X^{\mu} \rightarrow X^{\mu} + a^{\mu}$
with a constant $a^{\mu}$, and has a discrete symmetry defined by
$X^{\mu} \rightarrow X^{\mu},P_{\mu} \rightarrow P_{\mu},
 {\cal C}^{\pm} \rightarrow {\cal C}^{\mp},
\overline{{\cal P}}_{\pm} \rightarrow \overline{{\cal P}}_{\mp},
 \dl \rightarrow - \dl$.
We therefore assume, without loss of generality,
that $\omega$ also
respects the same symmetries\footnote{The algebraic equation (11)
to be solved is completely gauge-independent. It is
the gauge fixing which may break
the manifest Lorentz covariance.}.
The translational invariance
forbids
 $X^{\mu}$ to appear in $\omega$ without derivatives.  It is
convenient
to introduce the variables
\be
Y^{\mu}_{\pm} &\equiv& (P \pm X^{\prime})^{\mu}\ , \\
\de Y^{\mu}_{\pm} & = & \pm \dl ({\cal C}^{\pm} Y^{\mu}_{\pm})\ .\nn
\ee
In the  appendix, we show that there is no candidate for
$\omega$ which satisfies $\de\omega=0$. But there is a
unique solution with $\de \omega=$ a total derivative:
\be
\Omega &=& \int d\sigma\omega\ =\
\int d\sigma(\ \omega_{0} + \omega_{1}\ )\ ,
\ee
where
\be
\omega_{0} &= & k\ [\ ({\cal C}^{+} \dl^{3} {\cal C}^{+}) -
  (+ \rightarrow - )\ ] \\
  {\omega}_1 &=& k^{\prime}
 \  [\ ({\cal C}^{+} \dl {\cal C}^{+} +{\cal C}^{-}
 \dl {\cal C}^{+}) - (+ \leftrightarrow -)\ ]\
 Y_+ \cdot Y_-\ .
\ee
The $k$ and $k^{\prime}$ are gauge-independent constants
which should be calculated in some gauge. The formal solution (26)
contains an unfamiliar term, $\omega_{1}$ given in (28),
which depends
on string coordinates. This term is algebraically allowed
as an independent anomaly, but it would have never appeared in explicit
computations.  Thus it implies that we may demand
\be
k^{\prime} &=&0\ .
\ee

Now we would like to discuss the second consistency condition (12).
Since the total hamiltonian depends on the gauge chosen,
the solution of (12) depends on it, too. The gauge-fixed action in the
BFV formalism is defined as
\be
S &=& \int d^{2}\sigma
[\ P\cdot\dot{X}+\overline{{\cal P}}_{A}
 \dot{{\cal C}}^{A}+\overline{{\cal C}}_{A}\dot{{\cal P}}^{A}+B_{A}
 \dot{N}^{A}\ ]-\int d\sigma^{0}\ H_{\rm T}\ ,
\ee
with
\be
H_{\rm T} &=&  H_{\rm C}+{1 \over {i \hbar}} [Q\ ,\ \Psi]\ ,
\ee
where $ H_{\rm C}$ is the canonical hamiltonian,
and $\Psi$ is the gauge fermion \cite{HenA}.  For the present case,
$ H_{\rm C} = 0$. Combining the consistency condition
(12) with (31), we find that
 $\Ga$ is given by a double commutator, to which  we apply a super-Jacobi
identity to find
\be
\Ga &=& \{\Omega\ ,\ \Psi\}\ .
\ee
Clearly, we can not go further without any assumption on $\Psi$, and so
we restrict ourselves to the standard form of the gauge fermion
\be
\Psi &=&\int d \sigma [\ \overline{{\cal C}}_{\rm A} {\chi}^{\rm A} +
\overline{{\cal P}}_{\rm A} N^{\rm A}\ ]\ ,
\ee
where ${\chi}^{\rm A} $'s are gauge-fixing functions.
We look for the solution again in the form
\be
\Ga = \int d\sigma \gamma ,
\ee
with $\mbox{gh}(\gamma) = 1$ and $\mbox{dim}(\gamma) = 3~+~c$.
Since $~\Omega$ with $k'=0$
contains only ${\cal C}^{\pm}$ (see (26) and (27)),
we can compute the naive
commutator in (32) unambiguously if $\chi$'s  in $\Psi$
do not depend on
$\overline{{\cal P}}_{\pm} $. We therefore assume this, and
arrive at the
unique solution
\be
\gamma &=&
2 k [\ (\dl N^+ {\dl}^2 {\cal C}^{+}) - (+ \rightarrow -)\ ]\ .
\ee
This result is independent of the gauge-fixing
functions ${\chi}$'s, as long as the above assumption
, which is about the weakest one imposed on $\Psi$,
 is satisfied.

Although $\Omega$
with $k'=0$  has exactly the same form as the
$Q^{2}$ anomaly of ref. \cite{KOA}, the theoretical content in $\Omega$ is
much richer.
This $\Omega$, along with its descendant $\Ga$,
exhibits namely the most general form of
anomaly in the
extended phase space $-$ we  would like to call it the {\em generalized
Virasoro anomaly} because it must originate from the anomalous commutators
for the generalized Virasoro constraints,
$~\Phi_{\pm} = \{ \overline{{\cal P}}_{\pm},~Q \}$ \cite{FujB}.
Moreover, the result is pregeometric because
it has been obtained without any reference to a two
dimensional metric.
Note that
the geometrical meaning of
${\la}^{\pm}$ and $\xi$, which is given in (14), has disappeared
in the extended phase space, as one can see from their BRST
transformations (21); they
are no longer related to some metric variables, since the associated ghosts,
${\cal C}_{\la}^{\pm}$ and ${\cal C}^{\xi}$,
are by no means the reparametrization ghosts and the Weyl ghost.
So at the present stage,
the generalized Virasoro anomaly can not be identified with the
reparametrization or Weyl anomaly;
to distinguish these anomalies from each other we certainly
need a metric.
In the next section,
we shall describe how to
construct the metric variables
out of ${\la}^{\pm}$ and $\xi$, and
how to geometrize the
generalized Virasoro anomaly.

\section{Geometrization to derive the Weyl anomaly}
The primary constraints
being proportional to $\pi^{\la}_{\pm}$ and $\pi_{\xi}$ ,
respectively ( see (14) ), are responsible for the fact
that the variables, ${\la}^{\pm}$ and $\xi$, have lost their original
geometrical meanings.
The aim of this section is to
geometrize the theory to express the
generalized Virasoro anomaly in terms
of the GL(2)-covariant variables. This will enable us to interpret
that pregeometric anomaly.

\subsection{$GL(2)$-covariant ghosts}
We begin by writing the gauge-fixed action defined in (30)
with the standard
form of $\Psi$ in (33) more explicitly:
\be
S &=& \int d^{2}\sigma\ \{\ P\cdot\dot{X}
+\pi^{\la}_{+}\dot{\la}^{+}+\pi^{\la}_{-}\dot{\la}^{-}
\pi_{\xi}\dot{\xi}
+\overline{{\cal P}}_{A}\dot{{\cal C}}^{A} \nn\\
& &-{\cal H}_{\rm CL}-{\cal H}_{\rm GF}
-{\cal H}_{\rm FP}\ \}\ ,\ (A={\la}^{\pm},\xi,\pm)
\ee
where
\footnote{We have suppressed here the Legendre terms,
${\overline{{\cal C}}}_{A} {\dot{{\cal P}}}^{A}
+B_{A} \dot{N}^{A}
= - \delta ({\overline{{\cal C}}}_{A} \dot{N}^{A})$, by
 shifting the gauge fermion, $\Psi \rightarrow \Psi
+ ({\overline{{\cal C}}}_{A} {\dot{N}}^{A})$.}
\be
{\cal H}_{\rm CL} &=&
 {1 \over 4} ( P +X^{\prime})^2\ N^{+}
+{1 \over 4} ( P- X^{\prime})^2\ N^{-} \nn\\
& &+\pi^{\la}_{+}N^{+}
+\pi^{\la}_{+}N^{+}
+\pi_{\xi}N^{\xi}\\
{\cal H}_{\rm GF} &=&B_{A}\chi^{A} \\
{\cal H}_{\rm FP} &=&
 \overline{{\cal C}}_{A}\ \de\chi^{A}
+\overline{{\cal P}}_{A}{\cal P}^{A}
+ [\ 2\overline{{\cal P}}_{+}
\dl {\cal C}^{+} +
\dl \overline{{\cal P}}_{+} {\cal C}^{+}\ ]\ N^{+} \nn\\
& &-[\ 2 \overline{{\cal P}}_{-} \dl {\cal C}^{-} + \dl
\overline{{\cal P}}_{-} {\cal C}^{-}\ ]\ N^{-}
\ee
Terms in the first line in ${\cal H}_{\rm CL}$ suggests that $N^{\pm}$
can be related to two of the metric variables.
Recalling that $P_{\mu}$ defined in the original lagrangian (13)
is given by
\be
P_{\mu} &=& -\sqrt{-g}g^{0\al}\ \dl_{\al}X_{\mu}\ ,
\ee
and rewriting the action as
\be
{\cal L} &=& P\cdot\dot{X}
- \frac{\sqrt{-g}}{2g_{11}}(\ P^{2}+(X')^{2}\ )
- \frac{g_{01}}{g_{11}}\ P\cdot X'\ ,
\ee
one can easily find such relations:
\be
\sqrt{-g}/g_{11} \sim N^{0}\ ,\ g_{10}/g_{11} \sim N^{1},
\ee
where
$N^{\pm} \equiv N^{0}\pm N^{1}$.
Note that the
l.h.s. of the expressions
in (42) are exactly those for ${\la}^{0}$ and ${\la}^{1}$ in (14).
Therefore, to recover
the
original meaning of ${\la}^{\pm}$, we use
two of the gauge
degrees of freedom ( there are five in the extended phase space ) to
impose two gauge conditions
\be
\chi_{\la}^{+} = {\la}^{+}-N^{+}~, \qquad
\chi_{\la}^{-} = {\la}^{-}-N^{-},
\ee
without changing the relation,
$\xi = \ln g_{11}$.
There are still three gauge degrees of freedom, which
we would like to regard as corresponding to
two reparametrization symmetries and one Weyl symmetry.
This is possible only if
we can go to a configuration
space which involves among others three independent
metric variables, $g_{\al\bt}$, along with two reparametrization
(anti-) ghosts, $C^{\al}
(\overline{C}_{\al}) $, and one Weyl (anti-) ghost,
$C_{W}(\overline{C}_{W})$, with the covariant BRST
transformations \cite{FujG}
\be
\delta\ g_{\alpha\beta} &=& C^{\gamma}\ \partial_{\gamma}
g_{\alpha\beta}+\partial_{\beta}C^{\gamma}\ g_{\alpha\gamma}
+\partial_{\alpha}C^{\gamma}\ g_{\beta\gamma}
+C_{W}\ g_{\al\beta}\nn \\
\delta\ C^{\alpha} & =& C^{\beta}\ \partial_{\beta}C^{\alpha} \\
\delta\ C_{W} & =&  C^{\alpha}\ \partial_{\alpha}C_{W}\ .\nn
\ee
We would like to construct these covariant ghost fields
from the BFV
ghosts fields, by using various equations of motion.
To this end, one has to specify the remaining gauge conditions.
  However, it is sufficient for us
to assume that the gauge-fixing functions $\chi^{\pm},~\chi^{\xi}$
 in the gauge fermion (33) do not depend
on $\overline{{\cal P}}_{A},
 \pi^{\la}_{\pm},$ and $\pi_{\xi}$.  They are arbitrary otherwise, and
 in this sense our analysis given below is still gauge independent.
The variations of $~\overline{{\cal C}}^{\la}_{\pm}~$
yield the equations of motion
for $~{\cal C}_{\la}^{\pm}~$:
\be
{\cal C}_{\la}^{\pm} &=& {\cal P}^{\pm}~,~
\ee
Similarly, the variations of
$~\overline{{\cal P}}_{\pm},~\overline{{\cal P}}^{\la}_{\pm}$
and those of $~{\pi}^{\la}_{\pm}$ and ${\pi}_{\xi}~$ give
\be
{\cal P}^{\pm} &=& \dot{{\cal C}}^{\pm} \pm
{\cal C}^{\pm}
{\buildrel \leftrightarrow \over \dl}
N^{\pm}\ \ , \nn\\
{\cal P}_{\la}^{\pm} &=& \dot{{\cal C}}_{\la}^{\pm}\ ,
\ {\cal P}^{\xi}\ =\ \dot{{\cal C}}^{\xi}\ ,
\ee
and
\be
N_{\la}^{\pm} &=& \dot{{\la}}^{\pm}\ ,\ N^{\xi}\ =\
\dot{\xi}
\ee
where $ {\cal P}^{\pm}={\cal P}^{0}\pm {\cal P}^{1}$ and
$A {\buildrel \leftrightarrow \over \dl} B = A \dl B - \dl A B$.

Note that, under the covariant BRST transformation (44),
$\tilde{g}_{00}\equiv g_{00}/g_{11}$ and $\tilde{g}_{01}\equiv g_{01}/g_{11}$
 are Weyl-invariant, i.e. $\de \tilde{g}_{00} $ and $\de \tilde{g}_{01} $
do not contain $C_{W}$, and
that these variables have been already expressed in terms of
${\la}^{\pm}$.
To find $C^{\al}$, therefore,
we compare
$\de \tilde{g}_{00} $ and $\de \tilde{g}_{01}$
with $\de {\la}^{\pm}~=~ {\cal C}_{\la}^{\pm}$
(see (21)) with the equations of motion (45) and (46).
One easily finds
\be
C^{0} = \frac{{\cal C}^{0}}{{\la}^{0}} ,\quad C^{1} =
{\cal C}^{1}- \frac{{\la}^{1}}{{\la}^{0}}~{\cal C}^{0} .
\ee
The relation between $~{\cal C}^{\xi}$ and $~C_{W}$ can be found
in a similar manner.  We compare the BRST
 transformation obtained from (44)
\be
\de \ln g_{11} &=& C_{W}+C^{\al}\ \dl_{\al}\xi
+2~C^{0\prime}~ \frac{g_{01}}{g_{11}} + 2~C^{1\prime}
\ee
with the one in terms of the BFV basis
\be
\de \xi = {\cal C}^{\xi}.
\ee
where we have used (45), (46) and
the gauge conditions (43).
We find
\be
  C_{W} &=& {\cal C}^{\xi} - V_{\cal C}^{+} + V_{\cal C}^{-}~,
\ee
where $~V_{\cal C}^{\pm}$ is defined by
\be
   V_{\cal C}^{\pm} = \frac{1}{2}~ G_{\pm} {\cal C}^{\pm}
                        +( {\cal C}^{\pm} )'
\ee
with
\be
   G_{\pm} =  \frac{1}{{\la}^{0}}~[~\pm  N^{\xi} + ({\la}^{0} \mp
{\la}^{1})~
\xi^{\prime} \mp 2~ {\la}^{1 \prime}~]
\ee
Eqs.(49) and (51) completely fix the relation between the ghosts
in the BFV basis and those in the GL(2)-covariant basis.

\subsection{Derivation of the Weyl anomaly}
We come to the central issue of the present work,
derivation of the Weyl anomaly from the generalized
Virasoro anomaly.
We denote the geometrized
$~Q^2$ anomaly as
$~{\Omega}_{g} = \int d \sigma {\omega}_{g}$ and its
descendant as $~{\Ga}_{g} = \int d \sigma {\gamma}_{g}$.
Two different expressions for $Q^2$, $~\Omega$ given in (26)
and $~{\Omega}_{g}$,
 should obviously belong to the
same cohomology class defined by the BRST transformation
 (21) in the extended phase space.  Therefore,
the difference between $\Omega$ and
${\Omega}_{g}$ should be a coboundary term:
\be
   \int d \sigma {\omega}_{g} = \int d \sigma {\omega}_{0} -
\int d \sigma k {\delta} {\eta}~ .
\ee
  It turns out
that the desired coboundary term
\footnote{We shall see below how to find it.}
is
given by
\be
    \eta =  U^{+} + U^{-} - \frac{1}{2} (G_{+} - G_{-}) {\cal C}^{\xi},
\ee
where
\be
    U^{\pm}=  \frac{1}{4}\ G_{\pm}^2 {\cal C}^{\pm}
            + G_{\pm} ( {\cal C}^{\pm} )'~.
\ee
To obtain the geometrized expression for ${\omega}_{g}$ ,
one replaces the BFV ghosts in (54) by the covariant ones after
performing the BRST transformation (21) on $\eta$.  Using the relations
\be
 V_{\cal C}^{\pm} &=& \frac{\partial U^{\pm} }{\partial G_{\pm}}~,  \nn \\
 \sqrt{-g} R &=& -( V_{N}^{+} + V_{N}^{-} )'
+ \frac{1}{2} (G_{+} - G_{-})\dot{}~, \nn \\
   V_{N}^{\pm} &=&  \frac{1}{2}\ G_{\pm} N^{\pm} + ( N^{\pm} )'~,
\ee
as well as the equations of motion (46) and (47),
 one indeed obtains the covariant expression
\be
{\omega}_{\rm g} =
k \ [\ \sqrt{-g} R C^0  C_{W} +  \sqrt{-g} g^{0 \alpha}
 C_{W} {\partial}_{\alpha}  C_{W}\ ] +\ \ \mbox{(total spatial derivative)}\ .
\ee

We now derive $~{\Gamma}_{g}$.
The local form of (32) with the phase-space expression of ${\omega}_{g}$
becomes
\be
    {\gamma}_{g} &=& \{ {\omega}_{g} , \Psi \}
            = \gamma + k\{ \{ Q , \eta \} , \Psi \}  \nn \\
    \ \ \        &=& \gamma + k~ \dot{\eta} - k~ \delta
                   ( \{ \eta  , \Psi \} )    \ee
where $\gamma$ is given in (35), and
we have used a super-Jacobi identity
and the equation of motion
$\dot{\eta} = \{ \eta , H_{T} \}$.
A little algebraic calculation with the identity
$~V_{N}^{+} - V_{N}^{-} =  N^{\xi} $ yields
\be
   {\gamma}_{g}  = -k \sqrt{-g} R C_{W} + ~~\mbox{(total spatial derivative)}~.
\ee
It should be remarked that
using the relation \cite{fik}
\be
 \delta (\sqrt{-g} R C_{W}) &=& {\partial}_{\alpha} I^{\alpha}~, \nn\\
    I^{\alpha}           &=& C^{\alpha} C_{W} \sqrt{-g} R
                                 + \sqrt{-g} g^{\alpha \beta}
                                   C_{W}{\partial}_{\beta} C_{W}~,
\ee
one can also confirm that the geometrized $~{\Omega}_{g}$
and $~{\Gamma}_{g}$ in (58) and (60)
satisfy the descent equation in the configuration space
 $\delta {\Gamma}_{g} = -
 {\dot{\Omega}}_{g}$ with $~I^{0} \propto {\omega}_{g}$
\footnote{Conversely, this suggests how to choose the coboundary term
(55).}. The ${\gamma}_{g}$ is just twice of the divergence
of the BRST current, $2~{\partial}_{\al} J^{\al} $, which
was explicitly calculated by Fujikawa \cite{FujG}
in the conformal gauge.
 Comparing (60) with Fujikawa's result one can fix
\be
k &=& -\frac{(D-26)}{24\pi}\ .
\ee

Given the geometrized form of the
generalized Virasoro anomaly expressed by (58) and (60),
its theoretical meaning becomes now more transparent.
We observe that the geometrized $~\Ga_{g}$ contains the Weyl ghost only
and there is no trivial
term  which can be added to $\Ga_{g}$ (like $\eta$ in (59))
to completely replace $C_{W}$ by the reparametrization ghosts, $C^{\al}$
\footnote{This should be compared with the case
of the chiral anomaly in chiral Yang-Mills theories in which the difference
between the axial and vector current anomalies is
BRST trivial.}.
We therefore conclude that, unless $D=26$,
the constraint corresponding to the Weyl symmetry is inconsistent
with the dynamics of the system and consequently it may not
be regarded as a constraint any more.
As the result, in the configuration space with
three independent metric variables,
the generalized Virasoro anomaly should be uniquely identified with
the Weyl anomaly.
This result on the Weyl anomaly,
according to the character of its derivation,
is non-perturbative
and independent of a choice of gauges and regularizations
(under minor assumptions).

It is worth finding the counter term in action needed to shift the
generalized Virasoro anomaly $\Omega$ of (21)into its geometrized form
 ${\Omega}_{g}$ of (58).  It can be calculated from the coboundary term (55)
, and is given by
\be
 {\cal L}_{\rm g} &=&
        - \frac{\hbar ~k}{4} ( V_{N}^{+} G_{-} +  V_{N}^{-} G_{+}
        + N^{+'} G_{+} + N^{-'} G_{-}) \nn\\
  &=& \frac{\hbar ~k}{4}
   \Bigl[{1\over\sqrt{-g}g_{11}}\Bigl(\dot g_{11}-2g_{01}^\prime
+{g_{01}g_{11}^\prime\over g_{11}}\Bigr)^2-{\sqrt{-g}\over g_{11}}
\Bigl\{\Bigl({g_{11}^\prime\over g_{11}}\Bigr)^2
-4\Bigl({g_{11}^\prime\over g_{11}}\Bigr)^\prime \Bigr\}\Bigr].
\ee

It may also be instructive to derive a gauge-fixed form of the
Weyl anomaly.
As an example, we consider the orthonormal gauge which
is given by
\be
{\chi}^{\pm} =~ N^{\pm}-1 ~, \quad
{\chi}^{\xi}  =~ {\xi} ~.
\ee
Using the naive equations of motion for the ghosts in this gauge
\be
C_{W} = - {\partial}_{\alpha} C^{\alpha}~ , \quad
{\partial}_{\mp} C^{\pm} = 0~,
\ee
we find from (58) and (60) that
\be
[Q~ ,~ Q] = ik ~ \int d\sigma \{ ~(
C^{+}~ {\partial}^3 C^{+})- (+ \rightarrow -)~ \} ~ ,
\ee
and
\be
[Q~ ,~ H_{\rm T}]  = 0~ .
\ee
This is the classical result of Kato and Ogawa \cite{KOA}, and corresponds to
the BRST version of the conventional Virasoro anomaly.
\footnote{One has to be careful in concluding that
(66) and (67) are the gauge-fixed form of the Weyl anomaly.
It is because that the same expressions can be obtained
as a gauge-fixed form of the
reparametrization anomaly in the Nambu-Goto string.
See \cite{mar} for a more detailed discussion on the similarities and
differences between the Polyakov string and the Nambu-Goto string.}

 We would like to remark, however, that
gauge-fixed
forms of an anomaly are a self-contradicting notion
\footnote{We are not allowed to gauge away the Weyl degree of freedom
in the presence of the Weyl anomaly, for instance.};
 it is not legitimate
to fix the gauge and then to use equations of motions
in the presence of the anomaly.
Note that in contrast to those "on shell" anomalies,
the generalized Virasoro anomaly is obtained here completely
off shell and the Weyl anomaly is derived from it by using only
the "safe" equations of motion which are not affected by the Weyl anomaly.

\section{Summary}
By applying the generalized hamiltonian method of Batalin, Fradkin
and Vilkovisky \cite{FVA},
we have quantized bosonic string theory to perform
an exhaustive algebraic analysis on anomalies in the extended phase space.
In doing so, we have obtained, without any reference to a
two dimensional metric,
the most general form of anomaly,
the generalized Virasoro anomaly which is expressed by
$\Omega$ and its descendant $\Ga$
given in
(26) and (35), respectively.
This pregeometrical
anomaly has been uniquely identified with the Weyl anomaly
in the configuration space, and at the same time we
have derived its most
general form, without assuming the weak gravitational field.
Our results are
non-perturbative, independent of
the regularizations and
the gauge-fixing
functions (under the assumptions specified there), and
valid for any D.

The absolute normalization for
the Weyl anomaly and hence for
the generalized Virasoro anomaly can be fixed
by an explicit computation.
We have used the result of Fujikawa
\cite{FujG} in the conformal gauge, and derived the
$Q^{2}$ anomaly in the orthonormal gauge,
which has been computed by Kato and
Ogawa \cite{KOA}. This is one of the examples for showing the hierarchical
relations among the anomalies in bosonic string theory;
In the unconstrained extended phase space,
the generalized Virasoro anomaly is
sitting on the top of the hierarchy of anomalies. And in its
subspace, in which the two dimensional metric variables can be identified,
this pregeometrical anomaly obtains its geometrical meaning
and appears as the Weyl anomaly.

We have found the local counter term which shifts the generalized Virasoro
 anomaly into the Weyl anomaly.
If one begins with the total lagrangian consisting of (13) and (63),
 one may be led to a reparametrization invariant but Weyl
 non-invariant theory.
It means that
the results obtained in this paper will be the starting point toward
 construction of subcritical string theory or 2 D gravity as an anomalous
 gauge theory.
  We shall discuss this issue in the forthcoming communications
\cite{FIKMT}.

\vspace{2cm}
We thank K. Fujikawa for careful reading of the manuscript
and instructive suggestions.
One of us (J.K) would like to acknowledge
the kind hospitality at the Universit{\"a}t Dortmund and at
the Max-Planck-Institut f{\"u}r
Physik in M{\"u}nchen.

\vspace{3cm}
\centerline{\bf\large Appendix}
\vskip .3cm
In this appendix we prove the uniqueness of the formal solution
for $\omega$ given in (27) with (28) and (29).
As discussed in sect. 3, $\omega$ is a function
of ${\cal C}^{\pm},Y^{\mu}_{\pm}
\equiv (P \pm X^{\prime})^{\mu},{\overline {\cal P}}_{\pm}$
and their derivatives only.
We start by writing
the most general form of $\omega$, allowed by
its ghost number, its canonical dimension
(see eq. (22)), and the rigid symmetries assumed there :
$$
 \omega = {\omega}_0 + {\omega}_Y + {\omega}_{\overline {\cal P}}\ ,
\eqno(A.1)$$
where
$$
 {\omega}_0 = k_0 [\ ({\cal C}^+ {\dl}^3 {\cal C}^+) -
(+ \rightarrow - )\ ]\eqno(A.2)$$
$$
 {\omega}_Y = [\ {\cal C}^+ \dl {\cal C}^+ (k_1 Y_+
  \cdot Y_- + k_2 Y^2_+ + k_3 Y^2_- )
+{\cal C}^+ \dl {\cal C}^- (k_4 Y_+\cdot Y_-$$
$$ + k_5 Y^2_+ + k_6 Y^2_- )\ ]
\  -[\ + \leftrightarrow -\ ]\eqno(A.3)$$
$$
 {\omega}_{\overline {\cal P}} =
[\ {\cal C}^+ {\cal C}^- {\dl}^2 {\cal C}^+ \ (k_7 {\overline {\cal P}}_+
 + k_8 {\overline {\cal P}}_-)
 + \dl {\cal C}^+ \dl {\cal C}^- {\cal C}^+ \ (k_9 {\overline {\cal P}}_+$$ $$
 + k_{10} {\overline {\cal P}}_-)\ ] + [\ + \leftrightarrow -\ ]\ .
\eqno(A.4)
$$
In eqs. (A.2)-(A.4), we have regarded total derivative terms as zero because
they do not contribute to $\Omega$.
Since the BRST variation of $\omega_{0}$ is already a total derivative, i.e.
$$ \de {\omega}_0 =
 - k_0 [\ \dl({\cal C}^+ \dl {\cal C}^+
{\dl}^2 {\cal C}^+ + (+ \rightarrow -)\ ]\ ,
\eqno(A.5)\ , $$
$\omega_{0}$ is clearly an independent part of $\omega$.

We next consider $\de {\omega}_{\overline {\cal P}} $, and find that
its $\overline
{\cal P}$- dependent terms can become a total derivative, namely
$$
\de {\omega}_{\overline {\cal P}}
  = - k_{10} \dl [\ {\cal C}^+ {\cal C}^-\
             \dl {\cal C}^+ \dl {\cal C}^-\
({\overline {\cal P}}_+ - {\overline {\cal P}}_-)\ ] +  (\
\mbox{terms without}\
 {\overline {\cal P}}_{\pm})\ ,
\eqno(A.6)
$$
if and only if
$$k_7 = 0, k_8 = k_9 + k_{10}\ .\eqno(A.7)$$
If these relations are satisfied, on the other hand,
${\omega}_{\overline {\cal P}}$ can basically
be absorbed into $\omega_{Y}$. To see this, one adds to
$\omega_{\overline {\cal P}}$
a BRST trivial term
$$ - [\ k_9\  \de ({\cal C}^+ \dl {\cal C}^- {\overline {\cal P}}_+) +
k_{10}\
        \de ({\cal C}^+ \dl {\cal C}^+ {\overline {\cal P}}_-)\ ] +
[\ +\leftrightarrow -\ ]\ ,
\eqno(A.8)$$
and then shifts $k_{3}$ and $k_{5}$ in $\omega_{Y}$ according to
$k_3 \rightarrow k_3 + {1 \over 4} k_{10}$ and $ k_5 \rightarrow
 k_5 + {1 \over 4} k_9	$, respectively.
Therefore, we can discard ${\omega}_{\overline {\cal P}}$ as an independent
solution.
The  BRST
transformation of the remaining term, $\omega_{Y}$,
becomes a total derivative
if and only if
$$k_3 = - k_5 = k_6\ ,\ k_1 = - k_4 \ . \eqno(A.9)$$
One indeed finds
$$
 \de {\omega}_Y = k_1 \dl[\ {\cal C}^+ {\cal C}^- \dl({\cal C}^+
 - {\cal C}^-) Y_+ \cdot Y_-\ ]$$
$$ +k_3 \dl[\ {\cal C}^+ {\cal C}^- \dl({\cal C}^+
+ {\cal C}^-) (Y^2_- - Y^2_+)\ ]\ , \eqno(A.10)
$$ if (A.9) is used.
Note however that terms
proportional to $k_2$ and those proportional to
$k_3 = - k_5 = k_6$ in $\omega_{Y}$
are coboundary terms ( up to
total derivatives ), because
$$
\de ({\cal C}^{\pm} Y^2_{\pm}) =
\mp {\cal C}^{\pm} \dl {\cal C}^{\pm} Y^2_{\pm}\eqno(A.11)$$ $$
\de ({\cal C}^{\pm} Y^2_{\mp}) = \pm \dl({\cal C}^+ {\cal C}^- Y^2_{\mp}) \pm
[({\cal C}^{\pm} \dl {\cal C}^{\pm} + {\cal C}^{\pm} \dl {\cal C}^{\mp}
+ {\cal C}^{\mp} \dl {\cal C}^{\pm})Y^2_{\mp}].
\eqno(A.12)$$
This shows that $\omega_{Y}$ with $k'=k_{1}$ and eq. (A.9) -
up to coboundary and total derivative terms -
is exactly $\omega_{1}$ given in eq. (28).

\end{document}